# A Hybrid Event Detection Approach for Non-Intrusive Load Monitoring

Mengqi Lu and Zuyi Li, *Senior Member, IEEE*

***Abstract*—Non-Intrusive Load Monitoring (NILM) is a practical method to provide appliance-level electricity consumption information. Event detection, as an important part of event-based NILM methods, has a direct impact on the accuracy of the ultimate load disaggregation results in the entire NILM framework. This paper presents a hybrid event detection approach for relatively complex household load datasets that include appliances with long transients, high fluctuations, and/or near-simultaneous actions. The proposed approach includes a base algorithm based on moving average change with time limit, and two auxiliary algorithms based on derivative analysis and filtering analysis. The structure, steps, and working principle of this approach are described in detail. The proposed approach does not require additional information about household appliances, nor does it require any training sets. Case studies on different datasets are conducted to evaluate the performance of the proposed approach in comparison with several existing approaches including log likelihood ratio detector with maxima (LLD-Max) approach, active window-based (AWB) approach, and generalized likelihood ratio (GLR) approach. Results show that the proposed approach works well in detecting events in complex household load datasets and performs better than the existing approaches.**

## I. Introduction

FOSSIL fuels are the largest energy sources for electricity generation in the U.S, about 62% of which are from natural gas and coal in 2018 [1]. An average of 29.0 billion cubic feet per day (BCF/d) natural gas was consumed for electricity generation in 2018 and is expected to increase by 1.3% in 2019 [2]. The large consumption of these energy sources and their increasing trend make energy conservation a challenging task. Meanwhile, a growing attention towards environmental issues, such as carbon dioxide emission reduction, global warming mitigation, and sustainable development, requires a reasonable way to manage energy wisely. Since 37.4% of the electricity were consumed by the residential sector, it is imperative to take actions to save energy in this sector. Some studies indicate that by providing the energy consumption information feedback to consumers will reduce energy waste [3].

In recent years, the widespread implementation of smart meters has made appliance load monitoring (ALM) a promising way to cut energy consumption. A meter-level system, namely non-intrusive load monitoring (NILM), is a practical method to provide appliance-level electricity consumption information to both customers and utilities. NILM system aims to monitor appliances switching-on and -off actions, their working durations, and other relevant information. Through these data, NILM is able to decompose overall power consumption into the power usage of individual appliances. Unlike intrusive load monitoring (ILM), the NILM technique uses simple hardware (one sensor per house) but complex software (more computing) to achieve its goal. In this way, less sensor is needed which has benefits of low cost, easy installation, removal and maintenance. Due to its unique characteristics, the NILM technique could be used in energy management, power security, abnormal action detection and demand response (DR) for utilities while assisting customers to understand bills, plan monthly budget, and save money.

Existing NILM methods can be roughly classified into event-based and non-event-based depending on the requirement of obtaining appliance events. The changes or discontinuities in power signal are called events or edges which are treated as state transitions points in NILM. Event-based method requires the process of event detection and classification in order to further identify the appliances. Fig. 1 illustrates the active power measurements of a household lighting system which is recorded in the lab. The state transition points of each appliance are labelled as events shown in Fig. 1.

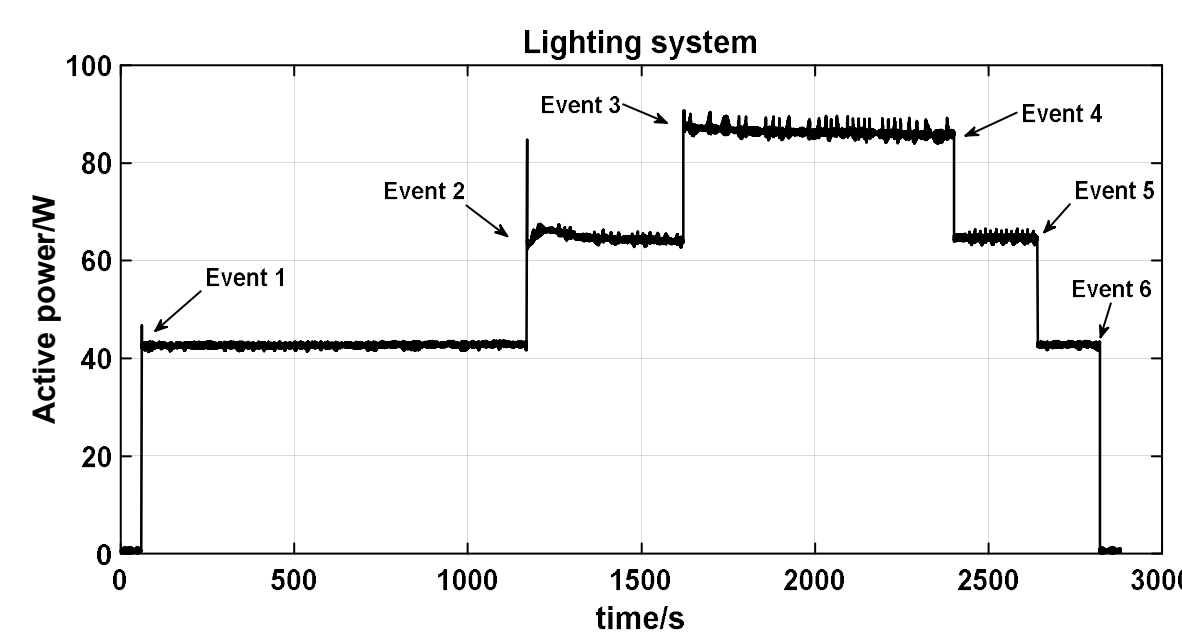

Fig. 1. The aggregated power and state transition events of a household lighting system measured at 20 Hz

The purpose of the event detection process in a NILM system is to detect the times when state transition actions occur from the aggregated measurements. The state transition actions normally include appliance turn-on, turn-off, speed adjustments, and function/mode changes. An accurate event detection approach is the prerequisite for precise load identification and valid power consumption estimation.

This paper proposes a hybrid approach for event detection in complex household load datasets. By using the hybrid approach, different categories of events can be captured at a reasonable accuracy. This approach mainly has following advantages:

- It requires only aggregated measurement information and does not require additional information such as historical data, and the number, categories, and power ratings of appliances.
- It can work with aggregated measurement data in different sampling rates.
- It improves the detection performance of events that are caused by appliances with long transients without triggering false alarms.
- It improves the detection performance of events that are caused by appliances with near-simultaneous actions and high fluctuations.

The rest of this paper is organized as follows. Section II presents and discusses existing event detection approaches. Section III proposes a hybrid approach to detect events. In Section IV, the proposed approach is tested and evaluated in different household load cases. Section V provides the conclusion.

## II. Background

This section reviews two categories of existing event detection approaches and evaluates the performance of two typical statistical event detection approaches.

### *A. Approaches requiring additional information*

Besides aggregated measurements, the first category of approaches requires additional information such as historical data, number of appliances in the house, power ratings and types of appliances. These approaches include factorial hidden Markov model (FHMM) [4], artificial neural network (ANN) [5], etc. They usually require training or extra library. This category of approaches usually has nice performance within the trained house, and it can achieve load identification together with event detection. However, these approaches have difficulties to detect new appliances and require the load information for new houses. In this paper, we mainly focus on the second category of approaches that does not require extra information as discussed in the next section.

### *B. Approaches without additional information*

The other category of approaches does not require extra information about the appliance and house. Instead, it relies on common knowledge of circuits and appliances to obtain a general solution to fit most situations. Two major groups in this category of approaches are rule-based and statistical approaches.

The rule-based approaches are mainly based on finding the times and sizes of all step-like changes from normalized power. In [6], a steady period is defined to be one of the minimum lengths in which the input does not vary by more than a specified tolerance. The difference between the averages across each period of change gives the step size. Such approach can also be used to detect transient patterns. Ref. [7] uses a similar approach with template called transient event detector to search for significantly varying segments using change-of-mean detector, denoted as v-sections. If the v-section of different devices does not overlap at the same time, then the transient class can be detected. The event detector employs two transversal filters as pattern discriminator. The first one compares recorded wave shape with template vector to identify event. The second filter checks the data magnitude to exclude noise. Ref. [8] presents an active window-based (AWB) approach for event detection. This approach utilizes power threshold with number of zeros in contiguous steady state portion to capture events.

The second group of approaches uses statistical method such as generalized likelihood ratio (GLR) and control chart. In [9], the authors utilize spectral envelope coefficients (Fourier coefficients) as load signatures along with a modified GLR approach to detect events. This approach is discussed in more detail in the next section. Ref. [10] applies a log likelihood ratio (LLR) test and voting window approach to detect events.

### *C. Performance evaluation of two existing event detection approaches*

In this section, two typical existing event detection approaches are presented in detail and evaluated by using lab-collected data. We evaluate their performances and analyze the deficiency before proposing our hybrid approach.

#### *a. Cumulative Sum (CUSUM) approach*

In this part, a sequential analysis approach, namely CUSUM, is applied to the aggregated data for event detection. The main purpose of this approach is to determine how the samples varies from their mean values. This approach calculates the mean value within a sliding window first, then computes the cumulative sum of the samples by using the following functions:

$$Mean_i = \frac{1}{n}\left(\sum_{i}^{i+n-1} x_j\right) \tag{1}$$

$$S_i = S_{i-1} + (x_i - Mean_i) \tag{2}$$

where $n$ is the number of samples within the sliding window; $Mean_i$ is the $i^{\text{th}}$ moving average value; $x_j$ is the $j^{\text{th}}$ selected electrical parameters (e.g., real power, reactive power, current, and power factors); $S_i$ is the $i^{\text{th}}$ cumulative sum value. Some references utilize different ways to determine CUSUM value. For example, Ref. [11] uses the following function.

$$S_i = S_{i-1} + (x_i - Mean_i)^2 \tag{3}$$

In one experiment, both the traditional CUSUM approach and the CUSUM approach in [11] are applied to the aggregated active power of a lighting system that includes two incandescent lights, two compact fluorescent lamp (CFL) lights, and an LED light. The aggregated active power and CUSUM results are showing Fig. 2.

The CUSUM approach was originally employed to find anomaly from signals. When it is applied to detect on-line appliance activities, there are several differences. The first is how to determine the mean value. Unlike segmentation of an existing complete speech signal, the usage duration of each appliance remains unknown to the system. Thus, the mean value of each segment cannot be precisely calculated. Some algorithms such as the moving average algorithm used above have to be applied to estimate the mean value. The second difference is that there are always multiple independent appliances involved in the electrical signal which may include changes in variance, correlations and spectral characteristics. This distinction causes the CUSUM results not to center at a single value but to center around different stable values.

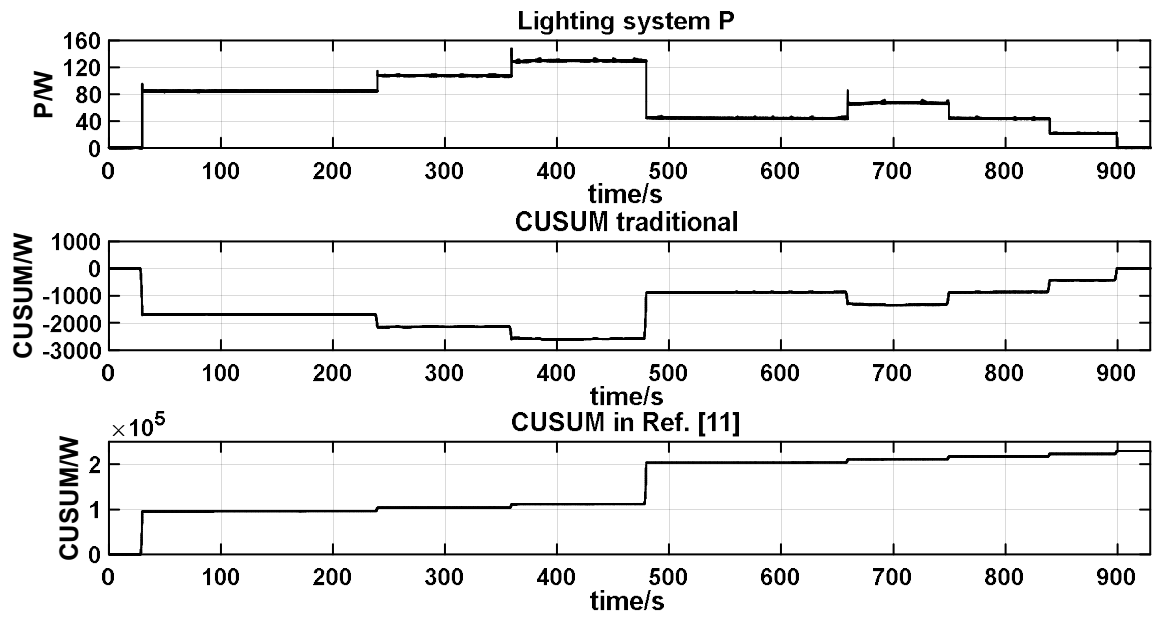

Fig. 2 The plots of active power and CUSUM results in both methods of a lighting system

However, the CUMSUM results still give indications to state transitions. In the plots of Fig. 2, the CUSUM value has a significant rise when power changes occur. Since the power changes for each appliance are different, the occurrence of such case makes it inappropriate to make a fixed upper bound and lower bound as is used for anomaly detection. Certain value differences [12] or other algorithms [13] are used to detect the power changes in the household datasets. The benefit of this approach is that it smooths the spikes at the turn-on point and oscillations during the working conditions. In order to detect the rise in the CUSUM, corresponding criterion to detect the rise section need to be applied. Ref. [13] provides a bootstrapping algorithm to use confidence level rather than setting threshold to detect possible events. The shortcoming of this approach is that it requires resampling 100 times or more for each segment of appliance actions.

*b. Log likelihood ratio approach*

Ref. [14] proposes a likelihood ratio-based approach, namely log likelihood ratio detector with maxima (LLD-Max). This approach relies on the change of mean value to calculate log likelihood ratio when the mean value of electrical parameters changes beyond specific value. The corresponding functions are defined as following.

$$ds(i) = \begin{cases} \frac{\mu_1 - \mu_0}{\sigma^2} \times \left| x_i - \frac{\mu_1 - \mu_0}{2} \right|, & |\mu_1 - \mu_0| > P_{th} \\ 0, & |\mu_1 - \mu_0| \le P_{th} \end{cases} \quad (4)$$

$$detection = max\left(abs\big(ds(i)\big)\right), \quad i - M_{pre} \le i \le i + M_{pre} \quad (5)$$

where $ds(i)$ is the $i^{th}$ detection statistics result; $x_i$ is the $i^{th}$ selected electrical parameter (e.g., real power, reactive power, current, and power factors); $\mu_0$ and $\mu_1$ are the mean values of $x_i$ within the preset window size before and after the current time, respectively; $P_{th}$ is the threshold of the selected electrical parameter; and $M_{pre}$ is the window size to find local peaks which is called maxima precision in the references.

This approach detects the event also based on the change of mean values. The calculation of detection statistics provides a way to find the location of changes. Figs. 3 and 4 show its performance on the same lighting system used in evaluating the CUSUM approach, as well as a more complex load dataset from a real office, respectively. The latter case includes three ceiling lights located in the living room, bedroom and bathroom, a water kettle, a toaster and a vent fan. Both datasets are collected at a sampling rate of 20Hz.

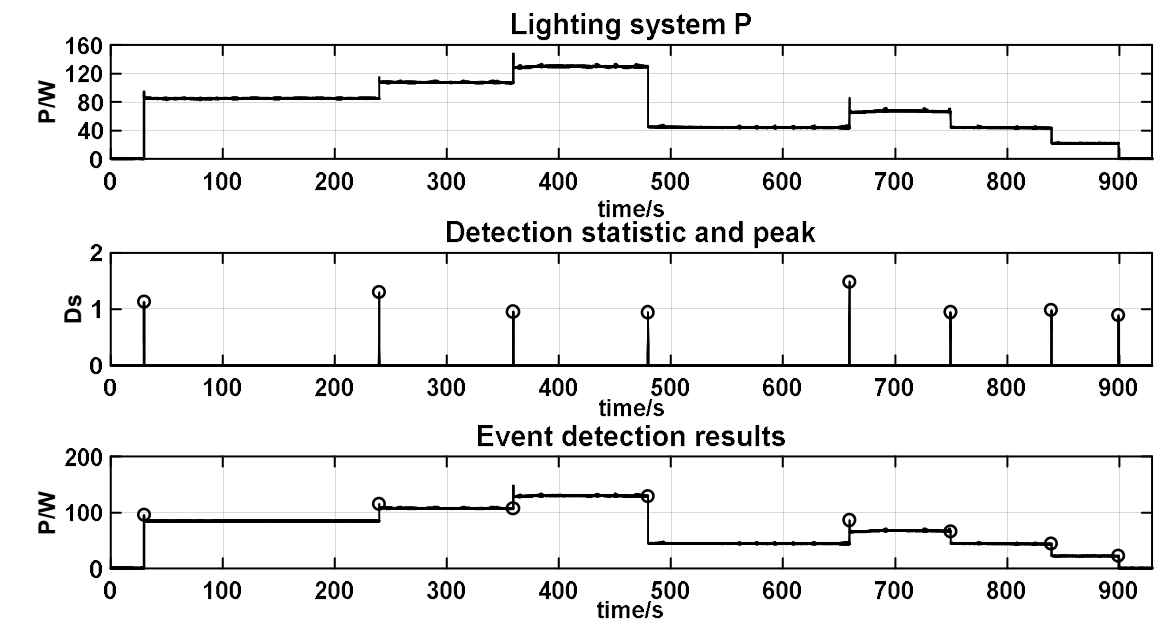

Fig. 3 The plots of active power and LLD-Max results in a lighting system

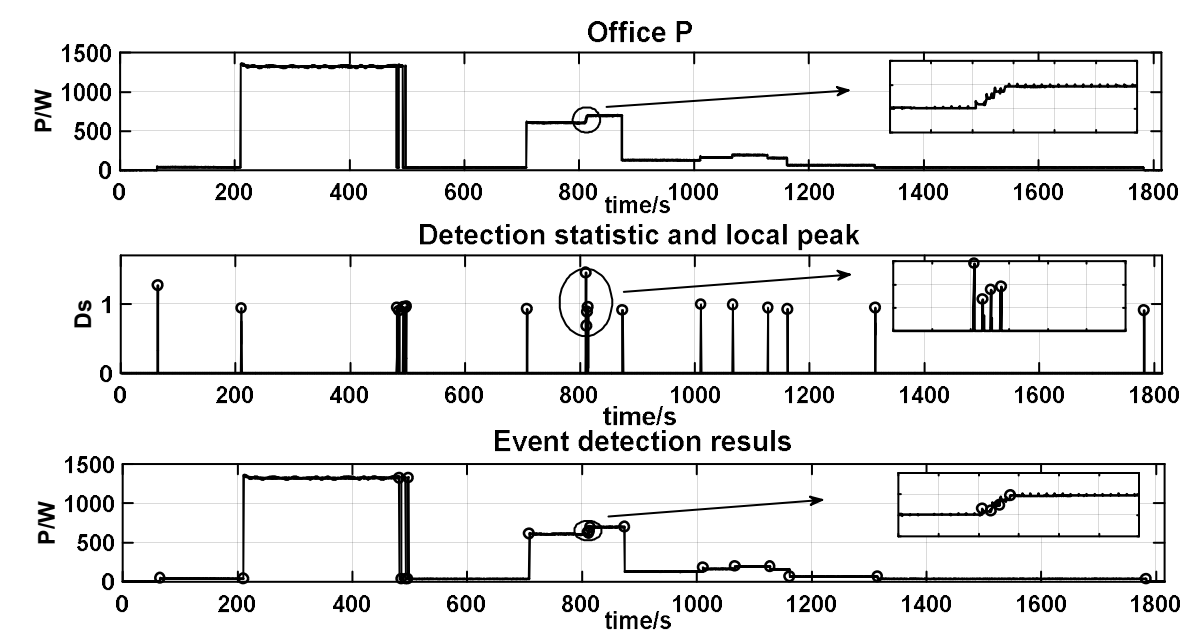

Fig. 4 The plots of active power and LLD-Max results in an office

As shown in Figs. 3 and 4, the LLD-Max approach detects all the events in both situations. However, it creates three false alarms when the ceiling light in the living room turns on. The reason is the long transient of that appliance.

Therefore, a new approach is required to deal with houses with complex appliance combinations such as the appliances with long-transient turn-on time or high fluctuations, user controlled continuous change devices like dimmable light and nearly simultaneous appliance actions with overlapping transients.

## III. Methodology

As described in the previous section, each event detection approach always has some deficiencies. In addition, the appliance characteristics in residential houses vary significantly such as difference in time and shape of transient states, range of fluctuations, and possible operation states. Thus, to come up with a single approach that works for a wide range of applications is a challenge since the electrical parameters of the aggregated measurements change whenever an appliance is connected to or disconnected from a household electric system. Our proposed approach pays more attention on the geometrical features of the aggregated data.

The proposed hybrid approach includes one base algorithm and two auxiliary algorithms aiming to detect true events and remove false ones. The selection of these algorithms is based on the analysis of typical false events and their causes. Fig. 5 depicts the flowchart of the proposed approach.

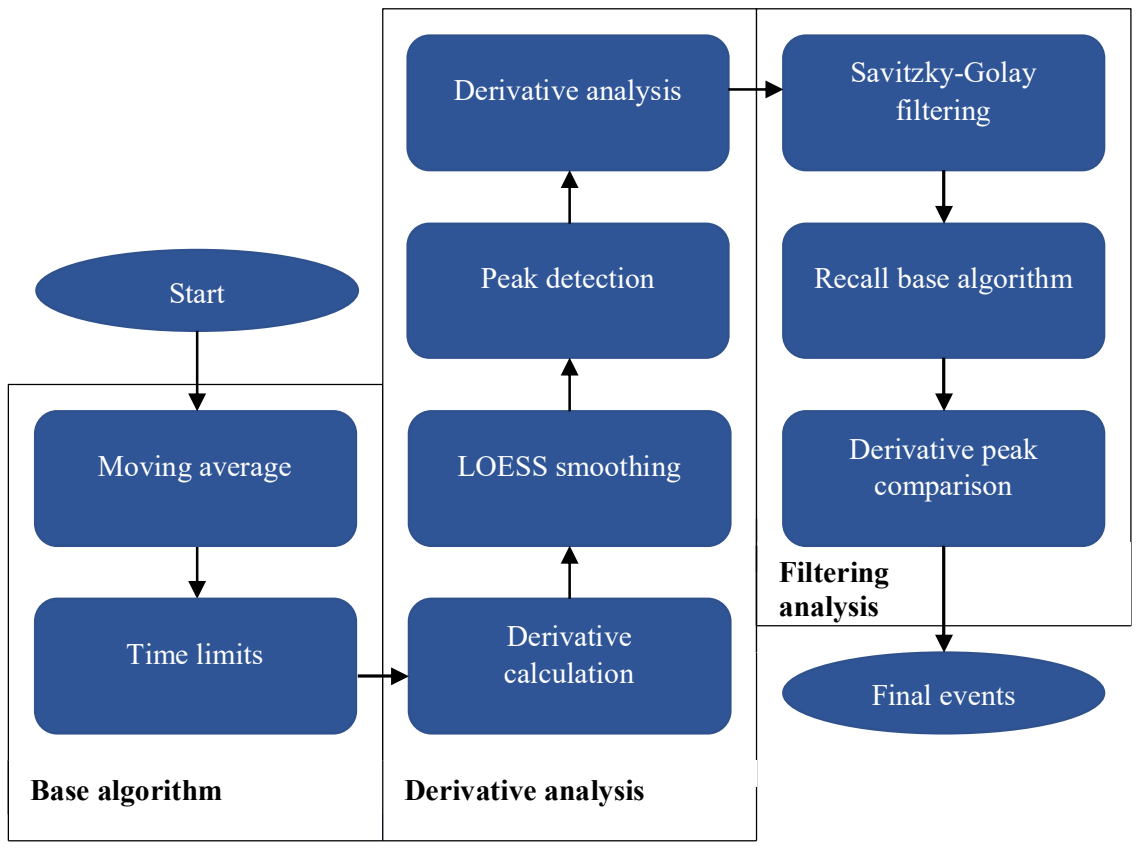

Fig. 5. Flowchart of proposed hybrid event detection approach

### A. Base algorithm – moving average change with time limit

In our work, we use moving average change with time limit as the base algorithm in the event detection process for the reasons listed below.

- This algorithm can capture most true events from the aggregated data.
- A small threshold of power change can be used as a fixed value in most situations.
- This algorithm is fast and easy to implement as an on-line method in NILM system.

This algorithm relies on the change of mean value for measured electrical parameters when appliance state transitions occur. In general, the mean values within a preset window are calculated once a new measurement collects from the aggregated data. When the difference between these two mean values is beyond certain threshold, an event is recorded. The functions are shown as follows.

$$Mean_{i_before} = \frac{1}{n}\left(\sum_{i-n-1}^{i-1} x_j\right) \tag{6}$$

$$Mean_{i_after} = \frac{1}{n}\left(\sum_{i+1}^{i+n+1} x_j\right) \tag{7}$$

$$\left|Mean_{i_{after}} - Mean_{i_{before}}\right| > P_{th} \tag{8}$$

where $n$ is the number of samples within the sliding window; $Mean_{i_before}$ and $Mean_{i_after}$ are the moving average values before and after the $i^{th}$ sample; $x_j$ is the $j^{th}$ selected electrical parameters (e.g. P, Q, I, and λ); $P_{th}$ is the threshold of electrical parameter. For the selection of $n$, we use a small value of samples within 0.3s. Large window size may lose the true events when two appliance activities occur within a short period of time. The other shortcoming associated with a large window size is the delay of detection time between the proposed base algorithm and actual measurement. Fig. 6 shows the detection results by using different window sizes. From the figure, significant rate of leading detection is appeared due to the large window size. The drawback of applying small window size is causing more false alarms which can be removed by using time limits and auxiliary algorithms in later process.

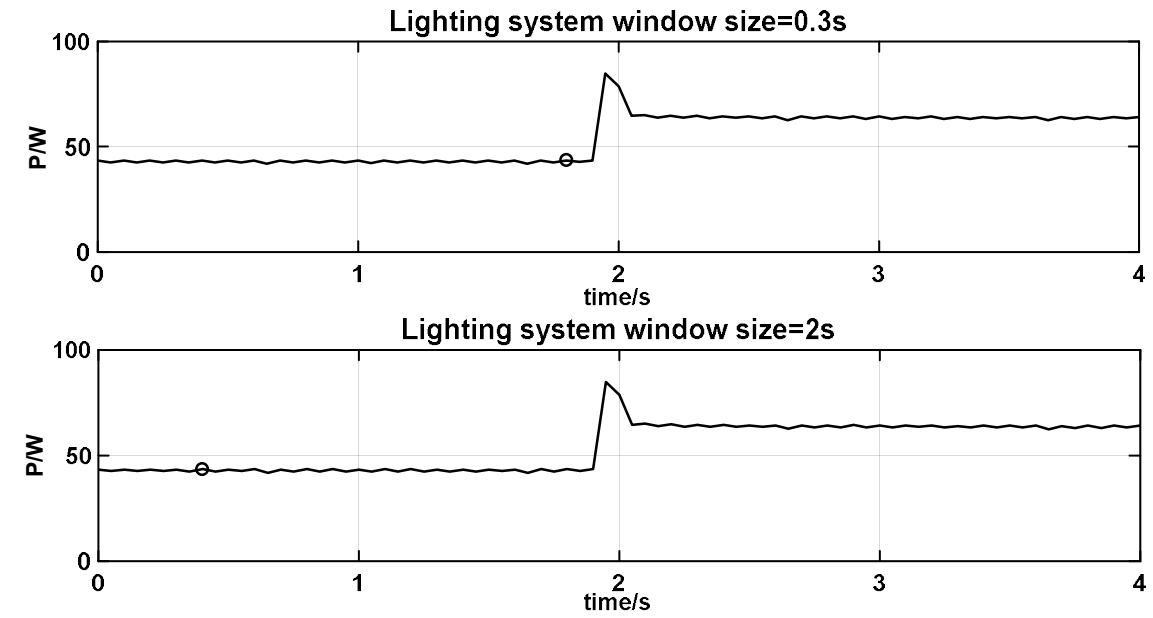

Fig. 6 The plots of event detection results by using different window sizes with and time limit in a lighting system

Some event detection methods utilize similar algorithm in change detection process since it is able to detect state transitions of appliances with significant distinct power values. The shortcoming of this method is the high false alarm rate and tuning of threshold values. The top plot of Fig. 7 shows the application of this algorithm in the lighting system collected in the lab. As the plot shown, all the events are detected but each event includes multiple false alarms. Therefore, the time difference between two continuous events is usually a limitation to alleviate this problem. Some methods use detection window size [14] which shares the same point of view. In the bottom plot of Fig. 7, the consecutive events within a threshold period are treated as a single event which further eliminates the false alarms from 24 to 0. This threshold or time limit is based on the survey of appliances transient characters in the house shown in Table I. In order to make the base algorithm more general to most cases, this time limit is set to 0.2s which means rapid power changes within 0.2s are treated as one event. We choose 0.2s since this value is long enough to separate the transients of most household appliances according to Table I. However, this method still triggers false alarms when there exists appliance with longer transient period than the selected threshold. The corresponding plots are shown in Fig. 8. The plots show that there are still multiple false alarms in the state transition of long transient appliance and high fluctuation device, however, the introduction of time limits reduce the amount of repeated false alarms.

The main limitation of this method is that it creates many false alarms if the threshold and window size are not properly selected. However, in order to develop a method which can be applied in various houses, these parameters are required as fixed values or common values in different situations. Therefore, the guideline of the proposed event detection method is to find all the possible events and then tries to remove these false alarms by auxiliary algorithms. Rather than using a single-step detector, sequential procedures are employed in this process. The motivation of design is based on the fact that the electrical behaviors of different types of appliance are difficult to be detected by a single algorithm.

The base algorithm in this paper utilizes power changes and time limits to detect events. We set the parameters of sliding window size, threshold of power changes and time limits as strict as possible so that this algorithm can be used as a general method for different datasets that does not miss true events as much as possible.

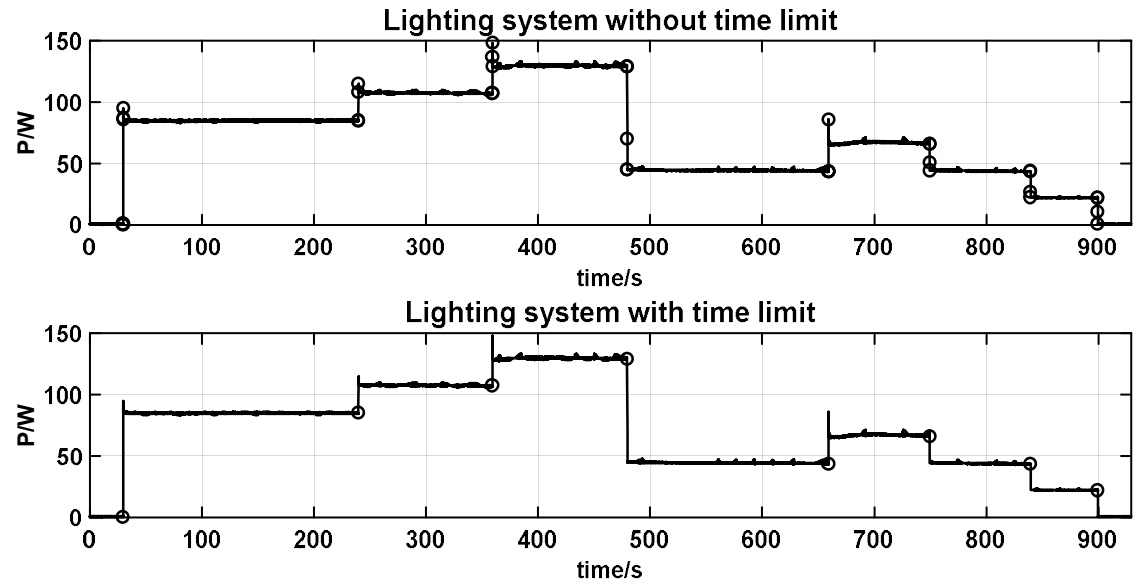

Fig. 7 The plots of event detection results by using moving average algorithm with and without time limit in a lighting system

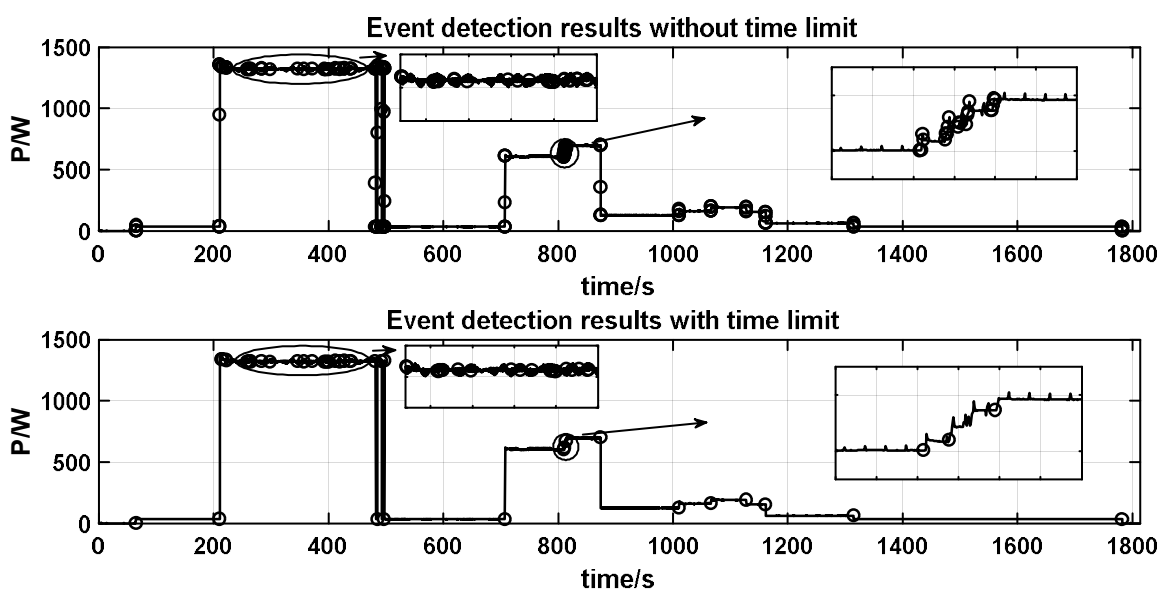

Fig. 8 The plots of event detection results by using moving average algorithm with and without time limit in an office

TABLE I
TYPICAL DURATIONS OF TRANSIENT PERIODS OF MAJOR APPLIANCES

| Appliance | Duration (s) |
|---|---|
| Air conditioner | 13 |
| Coffee maker | 0.1 |
| CFL | 0.2 |
| Fluorescent | 0.15 |
| Hair dryer | 0.2 |
| Electric kettle | 0.1 |
| LED | 0.12 |
| Refrigerator | 1.4 |
| Range hood | 2.8 |
| Toaster | 0.1 |
| TV | 0.3 |

The data of individual appliances is collected at the lab by using Yokogawa WT-3000E precise power analyzer. The measurements are recorded at 20 Hz.

### B. *Auxiliary algorithms*

From the experiments, we discover that most false alarms are caused by two main reasons, namely long transition time and high fluctuation. Therefore, we combine additional algorithms to solve these problems.

*a. Derivative analysis to handle appliances with long and complex transitions*

When the appliance changes its working states either turning on/off or switching functions, its electrical behavior changes and further results in the shifting of aggregated measurements. From Table I, the time limitation of two events is set to 0.2s which is long enough to separate state transition and different events for most household devices in the experiment. However, some devices have longer state transition time especially for turn-on transient period as shown in Table I and Fig. 9. In this case, the continuous changes in active power will keep trigger alarms until the device reaching steady state.

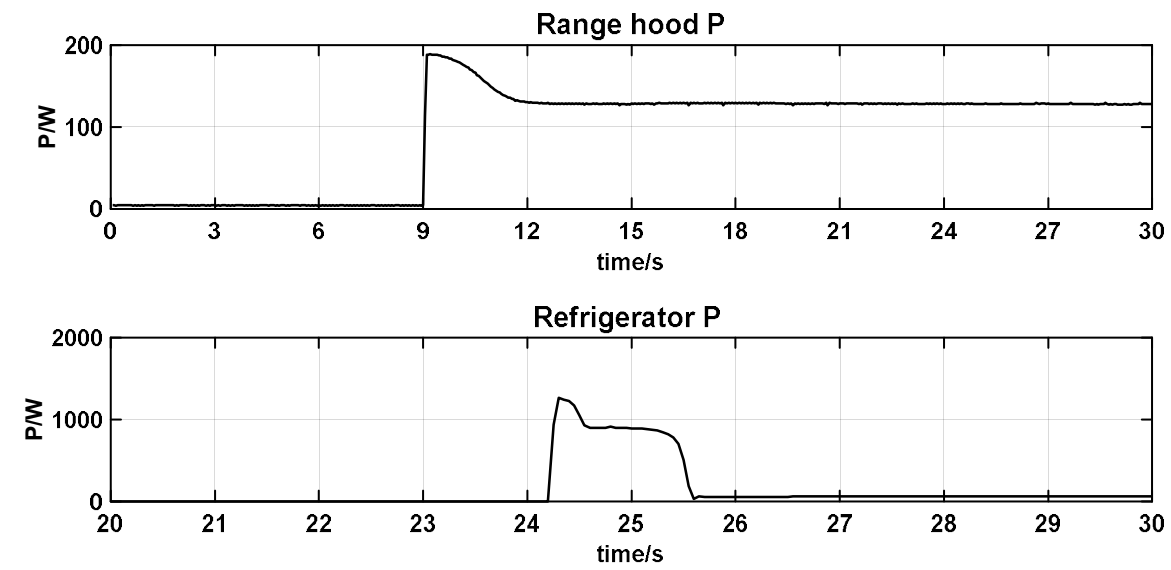

Fig. 9 The turn-on transient of range hood and refrigerator

In order to remove this type of false alarms, the shape of typical transient period needs to be carefully reviewed. As mentioned before, the purpose of event detection process in NILM is to detect state transition action which includes appliance turn-on, turn-off and function changes. The first plots of each row in Figs. 10 and 11 describe two typical turn-on transient waveforms of active power that we have observed in the experiment based on the measurements sampled at 20Hz.

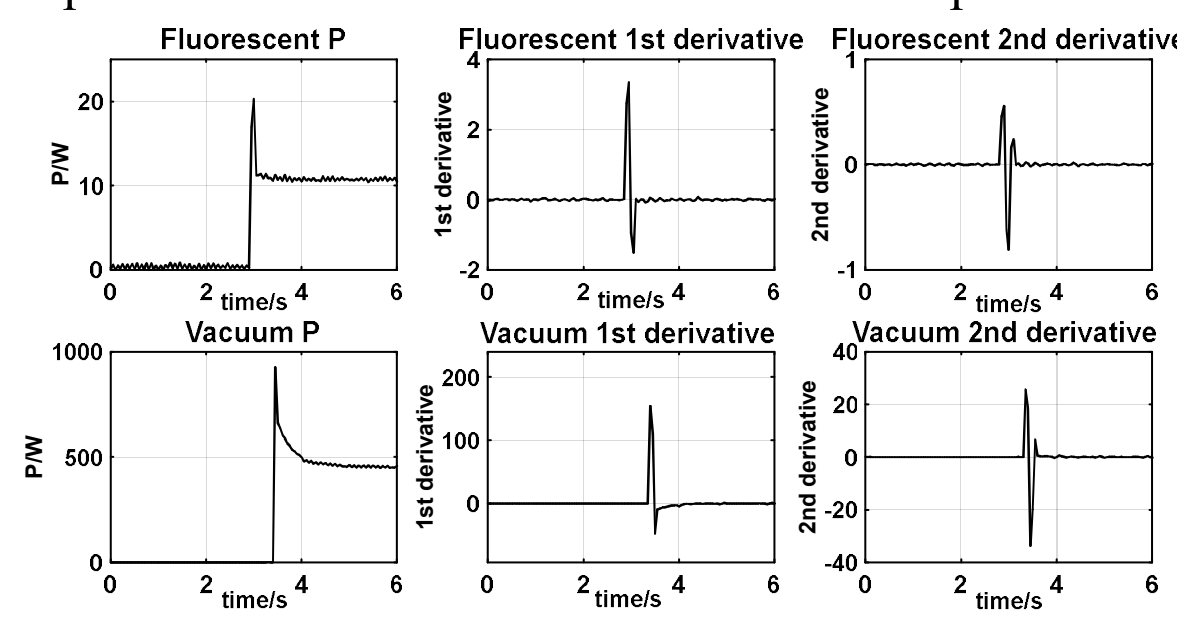

Fig. 10 The turn-on transient and their derivatives of type 1 appliances

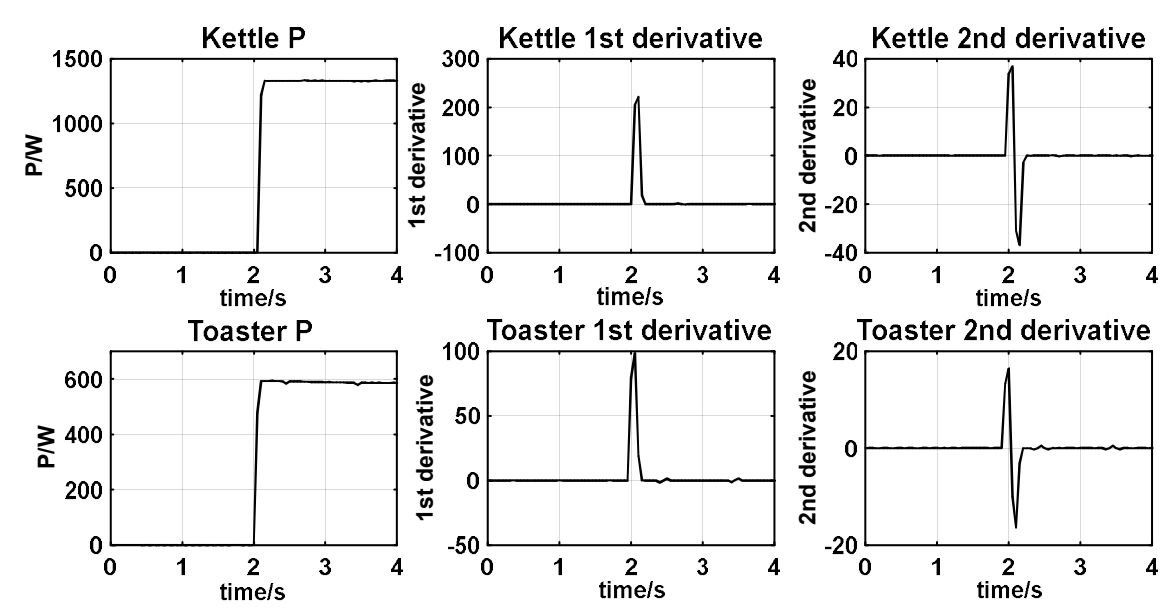

Fig. 11 The turn-on transient and their derivatives of type 2 appliances

The first type of devices has a spike at first then decline to steady state. Most motor-based appliances and lighting device such as blender, vacuum, CFL have this character. Turn-on transient of the other type of appliance is rising to steady state directly without showing significant spikes. Such appliances mainly include devices with heating elements such as electric kettle and toaster. The third type of appliance has more complex turn-on transient which may contain multiple spikes or different processes.

Different from turn-on period, turn-off transient waveform has only one power change pattern showing in the experiment which is simply decreasing to zero without any special features in active power. Fig. 12 illustrates this period.

Speed adjustment actions for appliance with motor and function changes of multi-function device usually occur during the working period of the appliance such as fan speed adjustment of the hair dryer and light operation installed in range hood. Fig. 13 presents one possible state transition pattern in the experiment which is different from the pattern of the above categories. The action shown in this figure is the fan speed adjustment of a range hood from high to low. The plot shows a negative peak (valleys) before the active power reaches steady-state value. Meanwhile, this period is a relatively long transient of 3 seconds.

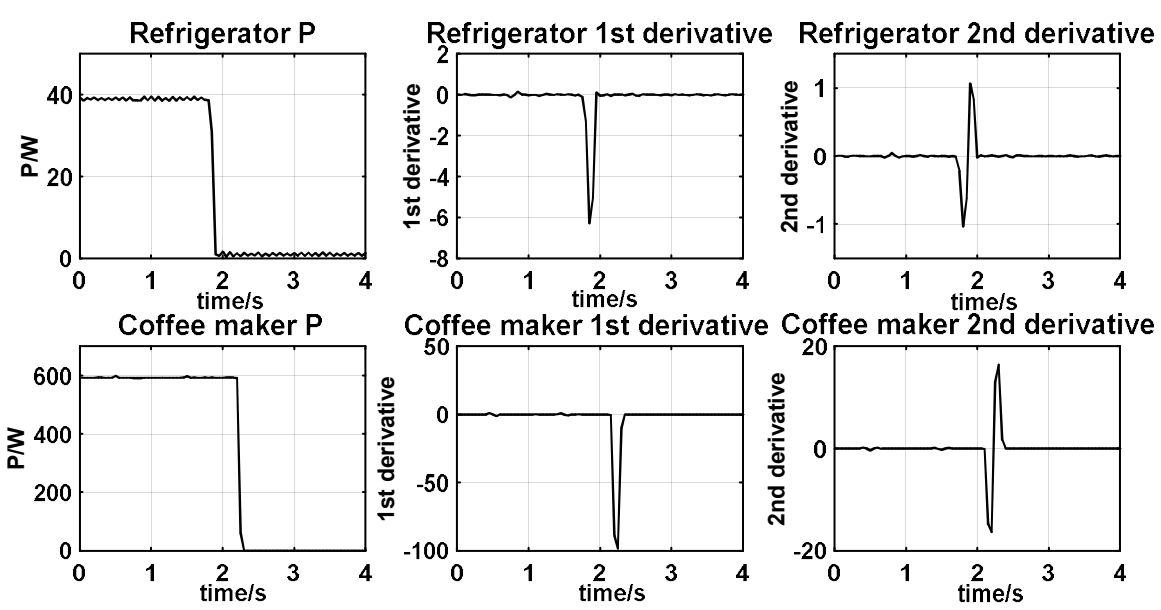

Fig. 12 The turn-off transient of appliances and their derivatives

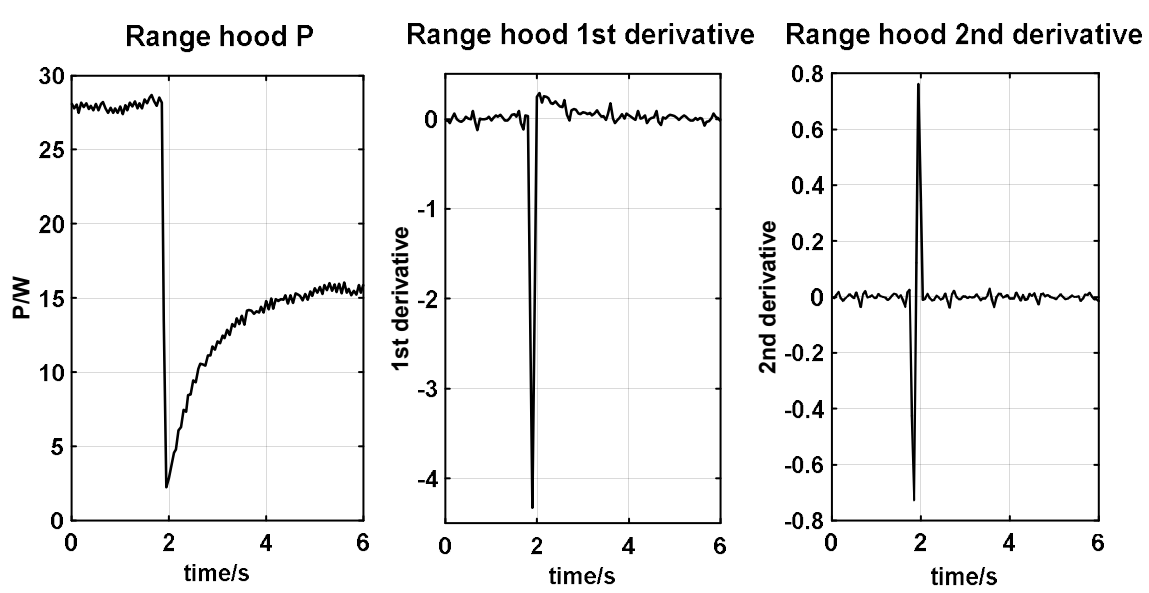

Fig. 13 The Speed adjustment transient of a range hood and its derivatives

For most household appliance, the steady-state is more stable and has a relative flat waveform. The transient-state, on the contrary, usually has relatively more changes and fluctuation. From the perspective of waveform shape, it is more likely to appear peaks and stages in the transient state. Similarly, appliance with long and complex transient also has these characteristics. Therefore, we use an auxiliary algorithm of derivatives to remove false alarms and to make sure that the event alarms once at the entire transient period. The first derivative describes the change direction and speed of the electric parameters which describes the slope of the aggregated data waveform. It depicts the increasing and decreasing of the appliance data. The second derivative is the derivative of the first derivative which describes the change speed of the 1st derivative. Both results show the shape of the transient period of appliance action. In addition, these data also can be signatures for feature extraction and load identification in later process. The functions of derivatives are listed as follows.

$$f' = \frac{x_j - x_{j-1}}{h} \tag{9}$$

$$f'' = (f')' = \frac{x_j - 2x_{j-1} + x_{j-2}}{h^2} \tag{10}$$

where $f'$ and $f''$ are the first-order 1st and second-order 2nd derivative of the electric parameter (e.g. P, Q, I, and $\lambda$) respectively; $x_j$ is the $j^{\text{th}}$ selected electrical parameters; and $h$ is the uniform spacing between two points. The 1st and 2nd derivatives of the turn-on transient for type 1 and 2 appliances are shown in the Figs. 10 and 11. The derivatives of turn-off and state change transient are shown in Figs. 12 and 13 respectively. In actual implementation, the 1st and 2nd derivatives can be used alone or as a combination. To make this paper clearer, we only use the 1st derivative in the rest of paper.

The derivatives like the original aggregated data contain noises which are rapid random changes. Thus, before applying peak detection process, a data smoothing method needs to be employed to reduce the noise so that the desired peaks will not be covered by these noises. In this process, we use locally estimated scatterplot smoothing (LOESS) [15] to achieve the goal. This algorithm uses weighted regression to smooth the data. It takes the neighborhood of the raw data using regression line to fit that data with certain weight and this line is considered as part of the smoothed data. Then the window slides for the next value. LOESS algorithm has the advantage of being flexible which is suitable to smooth complex data like household electrical measurements.

It should be noted that smoothing is not a way to eliminate the noise but to reduce the variance of the data. In this process, the main uncertainty is the selection of the slide window. However, since the purpose of this whole auxiliary method is to remove the long and complex transient false alarms, this period should be larger than the time threshold in base algorithm. We can determine this widow size according to the time limits as set in the base algorithm and the duration of long transient periods of major appliances in most houses like those in Table I.

After the smoothing process, the algorithm will detect peaks and valleys in the derivatives. Peak is a local maximum which is not smaller than its neighbors and valley is local minimum which is not larger than its neighbors as shown in (11). After smoothing, the detection of change point will turn to detect the significant peaks and valleys in derivatives. Fig. 14 shows the turn-on period of a range hood, 1st derivative and peak detection results of this long turn-on transient from top to bottom.

$$\begin{cases} peak\text{:}\ \ X(t) > X(t-1) \cap X(t) > X(t+1) \\ valley\text{:}\ \ X(t) < X(t-1) \cap X(t) < X(t+1) \end{cases} \tag{11}$$

where $X(t)$ is the time-series data at time $t$.

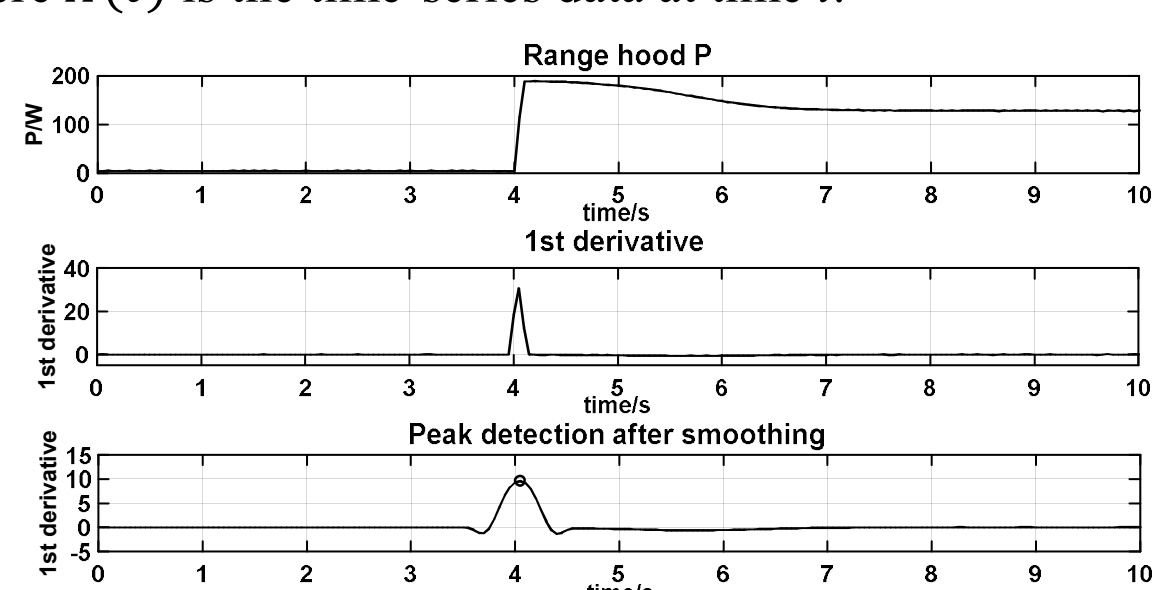

Fig. 14. Peak detection result after smoothing

Some appliances with long-transient and complex transition period may contain multiple events at their state transition period after peaks and valleys have been detected. The first and second plots of Fig. 15 present an example of refrigerator turn-on transient and its peak detection results. There are 6 peaks and valleys being detected in this period. In our algorithm, we define a true event as follows. In a state transition period, the absolute value of the 1st derivative will remain less than a small threshold for a relative long period. This definition is based on the principle that the magnitude of state transition will remain change all the time which can be expressed as non-zero value in the 1st derivative. When the appliance reaches its steady-state which means the data will keep at a fix value which can be expressed as a value around zero in 1st derivative. Since the noise exists, a small value is used instead of zero. Therefore, when the 1st derivative of electrical parameters (active power, current, etc.) falls into a near-zero value for a certain period we think this appliance reaches its steady state. In this way, the peaks and valleys in the middle can be treated as a single event. Thus, the false alarms have been removed. This process is described as follows.

$$d < \varepsilon \land t > Th \qquad (12)$$

where *d* is the 1st derivative of the aggregated data; $\varepsilon$ is a small preset value which is set to around 0.5 to get rid of the effect of noise; *t* is the time between two consecutive peaks or valleys; *Th* is the time threshold for an event which is set to 2s or longer according to Table I in our case. If (12) is satisfied, then the multiple events identified within the period *t* is regarded as one event; otherwise, these events are regarded as separate true events. The third plot of Fig. 15 shows the results after derivative analysis process. From the figure, the derivative analysis reduces 6 events after peak detection to 1 true event at the beginning of the turn-on transition.

The derivative analysis algorithm aims to remove false alarms caused by appliances with long and complex transitions. This algorithm uses derivatives of the aggregated data to identify multiple events locating at the same transition period. The derivate window size is usually set to 1 as unit spacing. We set the threshold to 0.5 in order to eliminate the interruptions of circuit noises. The LOESS window size and time threshold of multiple events are determined from the durations of long transient periods of major appliances in most houses.

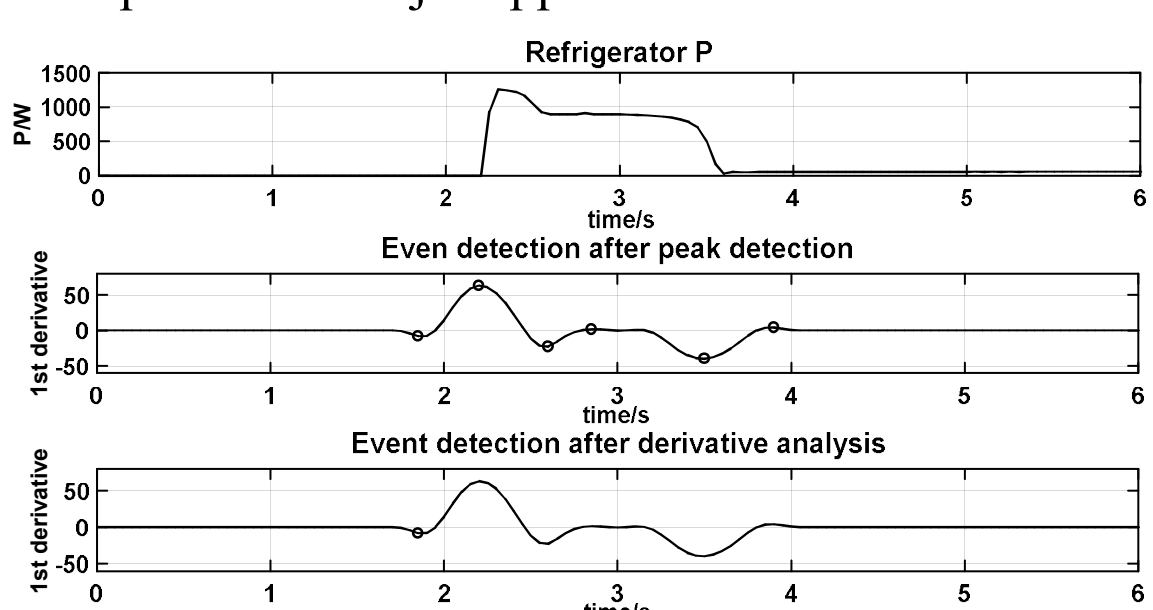

Fig. 15. Results of derivative analysis process

*b. Filtering analysis to handle appliances with high fluctuations*

As mentioned in the previous section, most household appliances work at nearly fixed power when they reach steady state. But some kinds of fluctuation exist due to different reasons such as its circuit structure, voltage oscillation, appliances interaction, etc. Small appliance usually creates less fluctuation which can be disregarded by the base algorithm. However, when multiple appliances are working at the same time or when a large appliance turns on, the fluctuation will reach a higher value which causes false alarms. Fig. 16 shows the event detection results when the base algorithm is applied to aggregated measurements of a kitchen. The circle in the bottom figure means the detected events. The tested appliance includes a coffee machine, a toaster and an electric water kettle and the detailed appliance activities are shown in Table II. When the electric kettle turns on, multiple events are detected even after the toaster turns off. In this case, the electric kettle enhances the oscillation of already-on appliance and further trigger alarms. The inherent way in the base algorithm to avoid this type of error is by using threshold and time of duration. Since both values are preset as fixed value and the number of working appliances in a house is assumed as unknown, additional filtering algorithm is required to solve this problem.

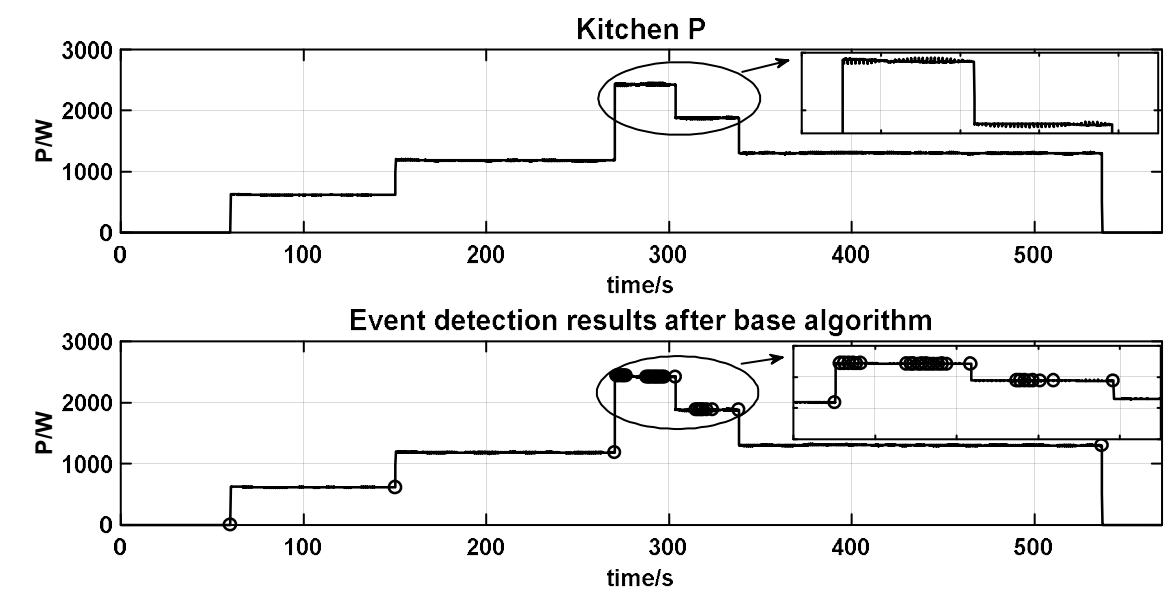

Fig. 16. Results of event detection by base algorithm

TABLE II
RECORD OF KITCHEN APPLIANCES IN THE EXPERIMENT

| State | Absolute time (hh:mm:ss) | Relative Time (s) |
|---|---|---|
| Start | 7:38:00 | 0 |
| Coffee machine ON | 7:39:00 | 60.1 |
| Toaster ON | 7:40:30 | 150.4 |
| Electric kettle ON | 7:42:30 | 270.3 |
| Toaster OFF | 7:43:03 | 303.55 |
| Coffee machine OFF | 7:43:38 | 338.4 |
| Electric kettle OFF | 7:46:57 | 536.95 |
| End | 7:47:30 | |
| Total time | 9.4983 min | 569.9 |

The first column shows the working state of each appliance. The second and third columns show the absolute time in the experiment and the corresponding relative time to the start of the experiment when appliance action occurs.

The filter can remove noise and unwanted fluctuations of the signal and leave the meaningful portion of the data. In the framework of NILM system, the filtering process are often followed by event detection algorithm to find the state transition of each appliance. Therefore, the task of filtering is to retain or enhance the state transition part while removing the undesired parts. However, during this process, the application of filtering often mistakenly removes true events or make events difficult to be captured by the base algorithm. Therefore, we choose to use filtering after base algorithm as an auxiliary method. The measurements can be expressed as following.

$$x_{meas} = x_{true} + e_{noise} \qquad (13)$$

where $x_{meas}$ and $x_{true}$ are measurement value and true value respectively; $e_{noise}$ is the noise or error in some publications. In the implementation, the true value here may not be the actual pattern of the appliance but a suitable form for precise detection by event detection method. In many cases, the selection of the filtering method depends on the event detection algorithm used in the system.

The low-pass filtering method has a close meaning of smoothing in this situation, which utilizes predicted values to replace measurements based on the adjacent points calculations. Therefore, a certain level of information loss will appear in the forms of event shape and amplitude distortion. Curve fitting applying polynomial function is a common and useful type of method to calculate prediction in above process. Such algorithms include Savitzky-Golay filtering, and mean filtering.

In our method, we use filtering after base algorithm to remove the false alarms triggered by high oscillation and noise. When the aggregated power beyond a preset threshold after a turn-on event occurs, the Savitzky-Golay filtering algorithm [16] is applied to the data. It fits higher order of polynomial to the original data by least-squares inside the window. This algorithm has advantage of separating components at different frequencies.

Even if the filtering is carefully applied, it still may remove true event. In order to make sure the removed point is a false alarm rather than true event, we use derivatives as restrictive condition to limit wrong elimination. In our algorithm, the removed events in the filtering process cannot be peaks and valleys in the derivative peak detection process. Fig. 17 shows the results of event detection in the kitchen after applying filtering algorithm. From the figure, all the false alarms are removed and only the true events remain.

The task of the filtering analysis is to remove the false alarms caused by appliances with high fluctuations. Savitzky-Golay filtering algorithm is applied to alleviate the effect of fluctuations in the data, then the base algorithm is run again to detect the events once more. In this way, some false alarms can be candidate events to be removed. This algorithm then compares these events with their derivative and removes the events whose derivative are not peak or valleys. This process ensures that the true events are not wrongly removed. During this process, the parameters used in the recalled base algorithm can be the same as those set previously or they can be stricter.

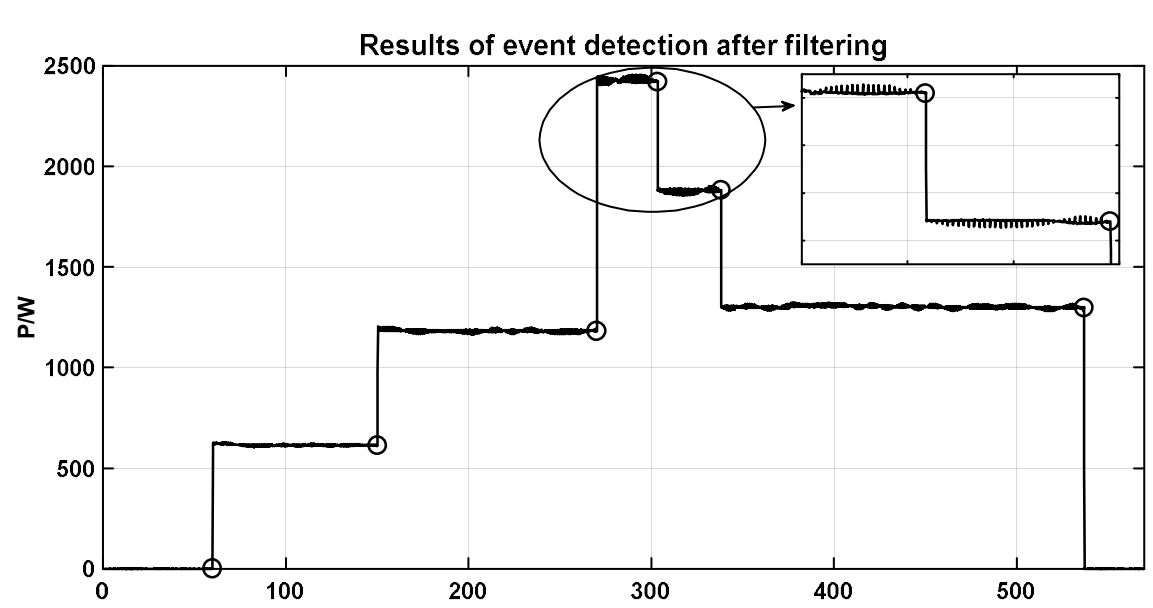

Fig. 17. Results of event detection after filtering process

## IV. Case Study

In this section, the proposed hybrid approach will be applied to various cases which aim to simulate different situations that may happen in real houses. The cases contain multiple types of appliances with different operation times. They are collected in real house, office and lab at standard voltage of 220V and frequency of 50Hz. We used Yokogawa WT-3000E precise power analyzer to collect the data at house level. The measurements are recorded at 20 Hz. In the last case, we use a one-day measurements in public datasets BLUED [17] to test our approach.

### *A. Household appliance including long transient and high fluctuation*

In this house, the tested appliances include an incandescent light, a toaster, a coffee maker, a range hood, and a hair dryer. The test lasted about 22.7 minutes with 20 appliance events including turn-on, turn-off, speed adjustment, and function switch. The detailed appliance events are shown in Table III and the aggregated data are shown in Fig. 18. The bottom two plots in Fig. 18 are the segments of aggregated data which show long turn-on transient period of range hood and high oscillation process of hair dryer with small appliance activity respectively. The purpose of this case is to test the performance of an event detection approach in the situations containing long transient appliances, high fluctuation appliances, and high fluctuation appliances with small power appliance actions.

TABLE III
Record of Appliances Activities in House 1

| State | Absolute time (hh:mm:ss) | Relative Time (s) | User Action |
|---|---|---|---|
| Start | 17:21:20 | 0 | |
| Kettle ON | 17:22:30 | 69.85 | |
| Incandescent light ON | 17:24:13 | 173.1 | |
| Toaster ON | 17:26:00 | 280.2 | |
| Range hood ON | 17:26:50 | 329.8 | Speed=3 |
| Kettle OFF | 17:27:35 | 375.2 | |
| Toaster OFF | 17:28:43 | 443.1 | |
| Range hood | 17:30:28 | 547.6 | Speed=2 |
| Coffee maker ON | 17:31:24 | 604.05 | |
| Range hood | 17:34:55 | 815.15 | Speed=1 |
| Incandescent light OFF | 17:36:00 | 879.9 | |
| Coffee maker OFF | 17:36:28 | 907.3 | |
| Range hood light ON | 17:36:44 | 924.1 | |
| Range hood light OFF | 17:37:53 | 992.3 | |
| Range hood OFF | 17:37:58 | 997.3 | |
| Hair dryer ON | 17:39:30 | 1089.3 | Heat=OFF |
| Hair dryer | 17:40:20 | 1139.25 | Heat=Low |
| Incandescent light ON | 17:41:30 | 1209.35 | |
| Hair dryer | 17:41:53 | 1232.45 | Heat=High |
| Hair dryer OFF | 17:42:27 | 1266.4 | |
| Incandescent light OFF | 17:43:16 | 1315.4 | |
| End | 17:44:00 | | |
| Total time | 22.7 min | 1359.4 | |

The last column shows the user action which includes fan speed control of range hood (1 to 3, 3 for highest) and heat mode of hair dryer (OFF, Low and High).

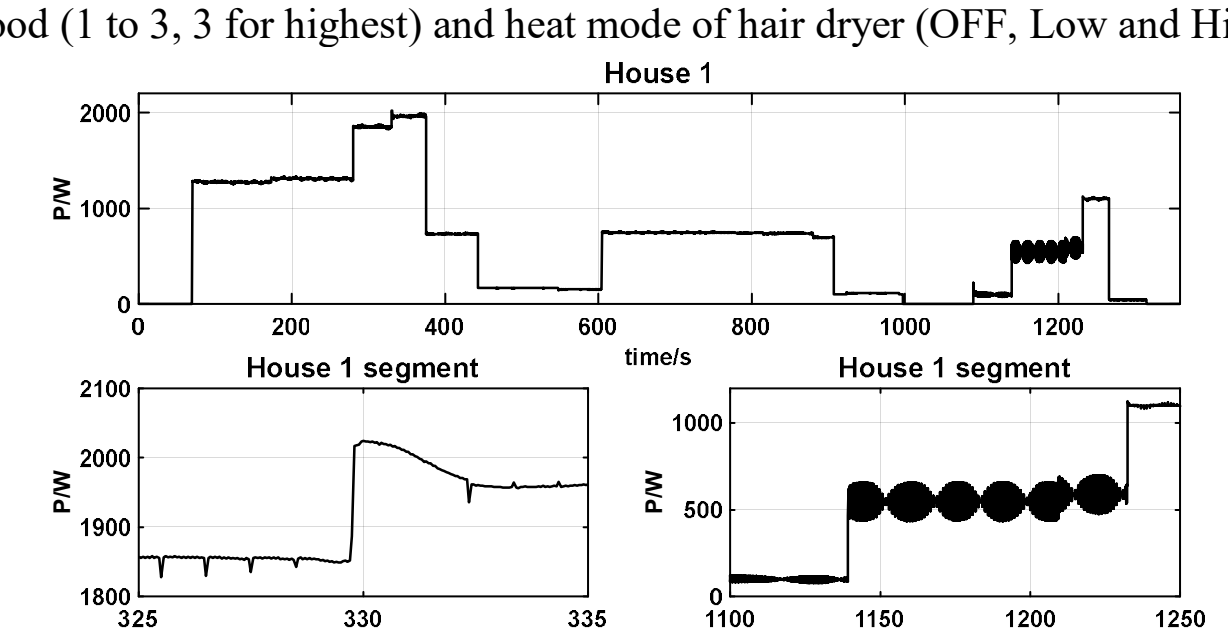

Fig. 18. Active power plots of house 1 measurements

Fig. 19 shows the event detection results by the base algorithm in the proposed hybrid approach. From the results, the base algorithm is able to find all the true events including turn-on action of an incandescent light during hair dryer working. The number of detected events is 31 whereas the number of true events is 20. The false alarms are caused by two special working processes of range hood and hair dryer.

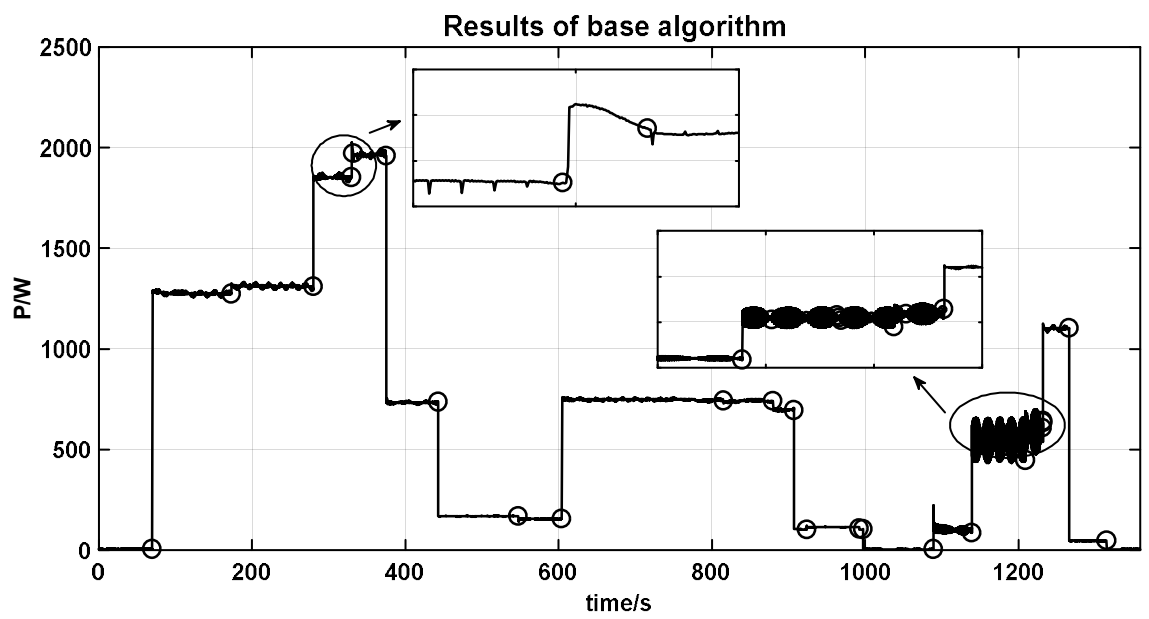

Fig. 19. Event detection results of base algorithm

Fig. 20 shows the event detection results of derivative analysis. The first plot from top depicts suspicious events which occur near each other. The second plot shows the derivative peak detection and analysis of these suspicious events. The algorithm finds two peaks in the above two periods which means there are only two true events among these alarms. In the last plot, these false alarms caused by complex transients are removed. The number of events is reduced to 22 since other false alarms are far from others which means they are not caused by long transients.

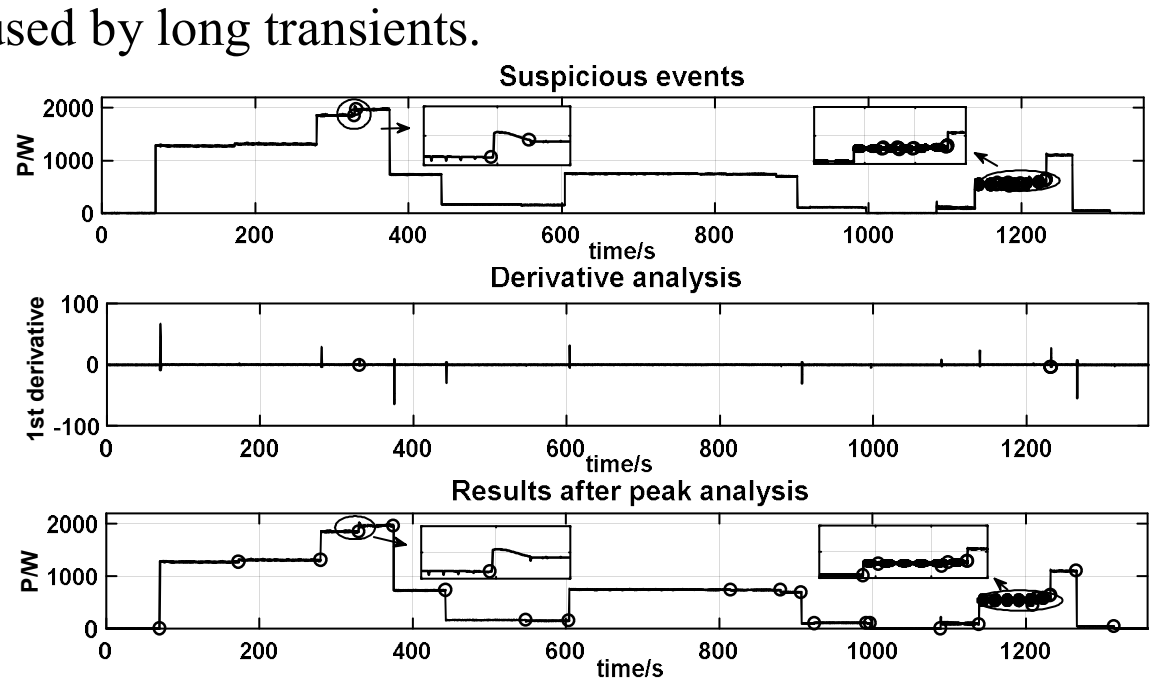

Fig. 20. Event detection results of derivative analysis

Fig. 21 shows the events detection results of filtering and final results. The first plot depicts the event detection results after filtering method applied to the original data. In this process, false events caused by oscillation are filtered by this algorithm. The removed events are described in the second plot. The last plot shows the final results which contains all the 20 true events without false event. Table IV records the number of event detection results after applying each algorithm.

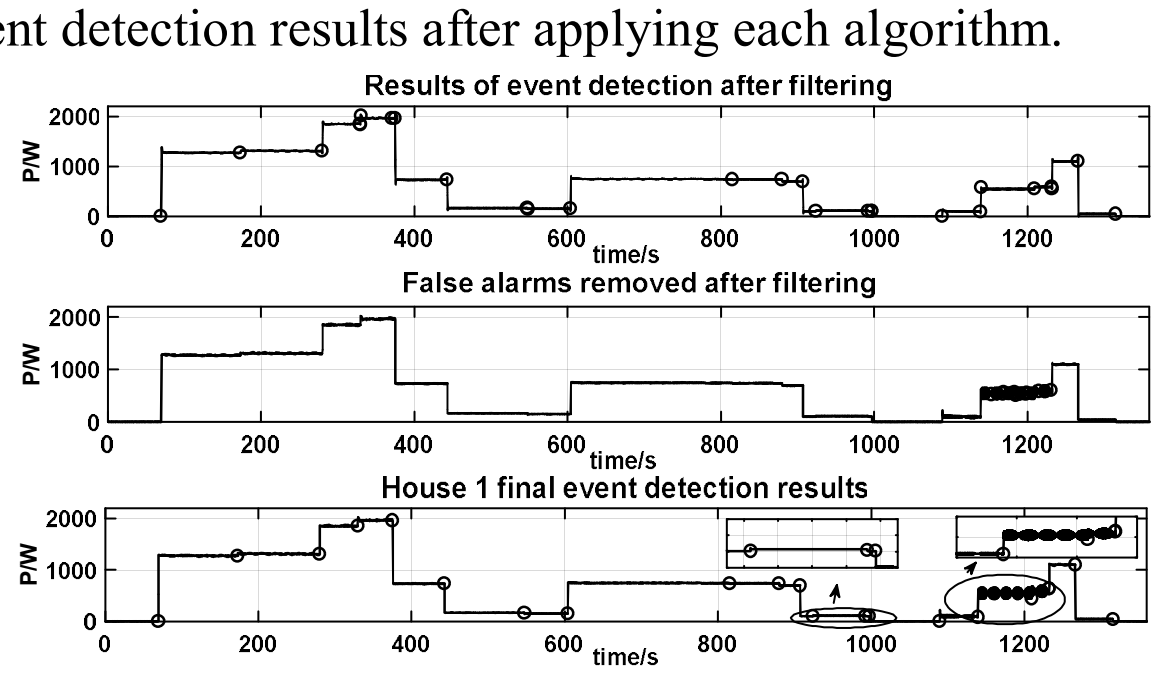

Fig. 21. Event detection results of filtering

TABLE IV
NUMBER OF EVENTS AFTER EACH ALGORITHM

| Algorithm | Events number |
|---|---|
| Base algorithm | 32 |
| Derivative analysis | 22 |
| filtering | 20 |

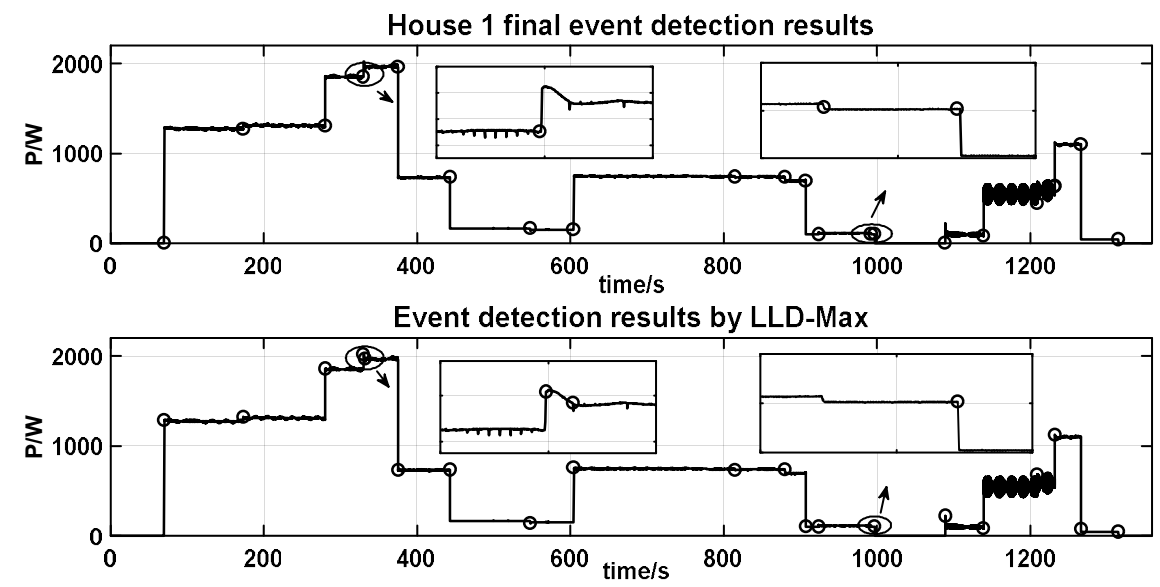

Fig. 22. Event detection results and comparison to LLD-Max algorithm

In this case, the proposed hybrid approach detects all the true events including some complex ones without causing false alarm. It proves that this approach has ability to capture events with complex transients and fluctuations. When small appliance has actions during the fluctuation power, this approach still has ability to detect them. In Fig. 22, we compare the detection results with LLD-Max algorithm [14]. From the results, the LLD-Max algorithm detects one false event and loses one true event results from long-transient and close-action.

### *B. Household appliance actions with short time interval*

This case contains lighting system of a residential house which includes two 40W incandescent lights, two 24W compact fluorescent lights (CFLs) and a 24W LED. Near simultaneous actions, which means multiple state changes occur within certain time (4s and less in this experiment), are taken in this experiment. The purpose of this combination is to test the performance of an event detection approach to near-simultaneous actions. The overall test lasted about 31 minutes and the detailed appliance events are shown in Table V. The corresponding plots and detection results are shown in Fig. 23.

TABLE V
RECORD OF APPLIANCES ACTIVITIES IN LIGHTING SYSTEM

| State | | Absolute time (hh:mm:ss) | Relative Time (s) |
|---|---|---|---|
| Start | | 18:11:00 | |
| CFL 1 ON | | 18:12:00 | 60 |
| CFL 1 OFF | | 18:15:00 | 240.3 |
| CFL 2 ON | | 18:15:30 | 270 |
| CFL 2 OFF | | 18:18:30 | 450.15 |
| Incandescent 1 ON | | 18:19:00 | 480.05 |
| Incandescent 1 OFF | | 18:22:00 | 660.2 |
| LED ON | | 18:22:30 | 690.2 |
| LED OFF | | 18:25:30 | 869.85 |
| Near-simultaneous | CFL 1 ON | 18:28:00 | 1020.15 |
| | Incandescent 1 ON | 18:28:04 | 1024.2 |
| Near-Simultaneous | LED ON | 18:30:30 | 1169.9 |
| | Incandescent 2 ON | 18:30:34 | 1173.8 |
| Near-Simultaneous | CFL 2 ON | 18:34:00 | 1380.05 |
| | LED OFF | 18:34:04 | 1384.2 |
| CFL 1 OFF | | 18:35:00 | 1439.75 |
| Incandescent 1 OFF | | 18:37:00 | 1559.9 |
| Near-Simultaneous | CFL 2 OFF | 18:40:00 | 1740.35 |
| | Incandescent 2 OFF | 18:40:01 | 1741.15 |
| End | | 18:42:00 | |
| Total time | | 30.9983 min | 1859.9 |

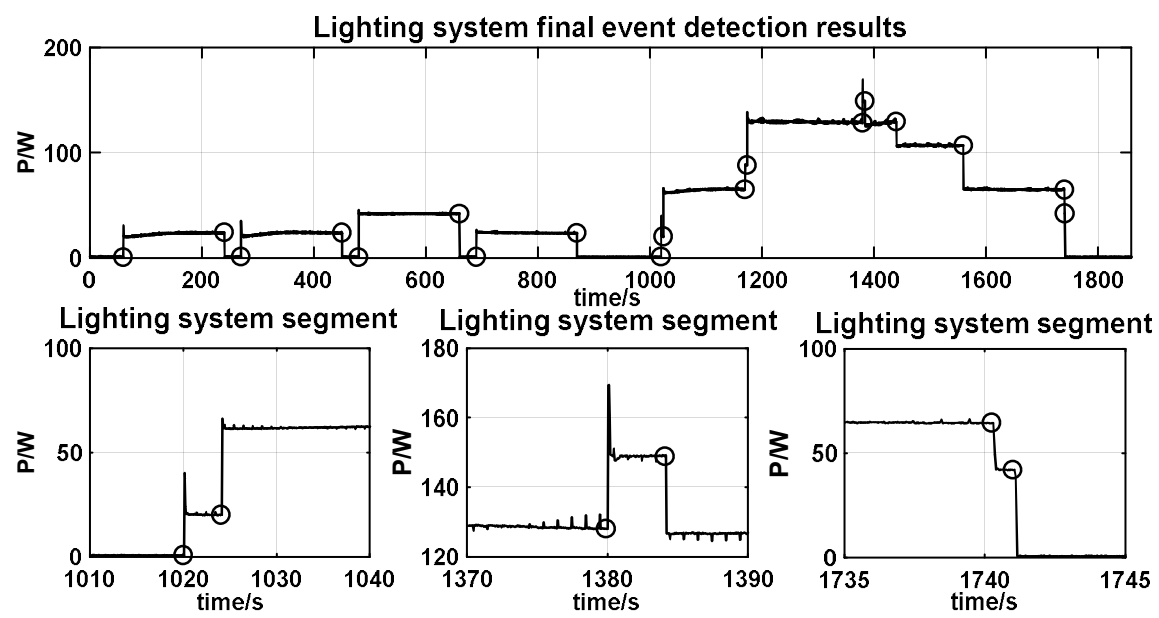

Fig. 23. Event detection results of a lighting system

In the detection results, all the 18 events are detected by the algorithm even when the time difference between two events is less than 2 seconds. In this case, these near-simultaneous events are mainly detected by the base algorithm. The derivative analysis focus on removing false alarms on the same transient. Since the active power in this lighting system is not high, there is no fluctuation caused false alarms. The filtering process does not remove events in this situation.

Fig. 24 also shows the event detection results by using AWB approach in [8]. Two plots in this figure illustrate the captured events with number of zeros = 1s and 0.1s from top to bottom. In Fig. 24, the AWB algorithm has 1 missing event if number of zeros sets to 1s. However, in order to detect the closed-action event, reduction of this parameter will cause one false alarm in relatively long-transient action.

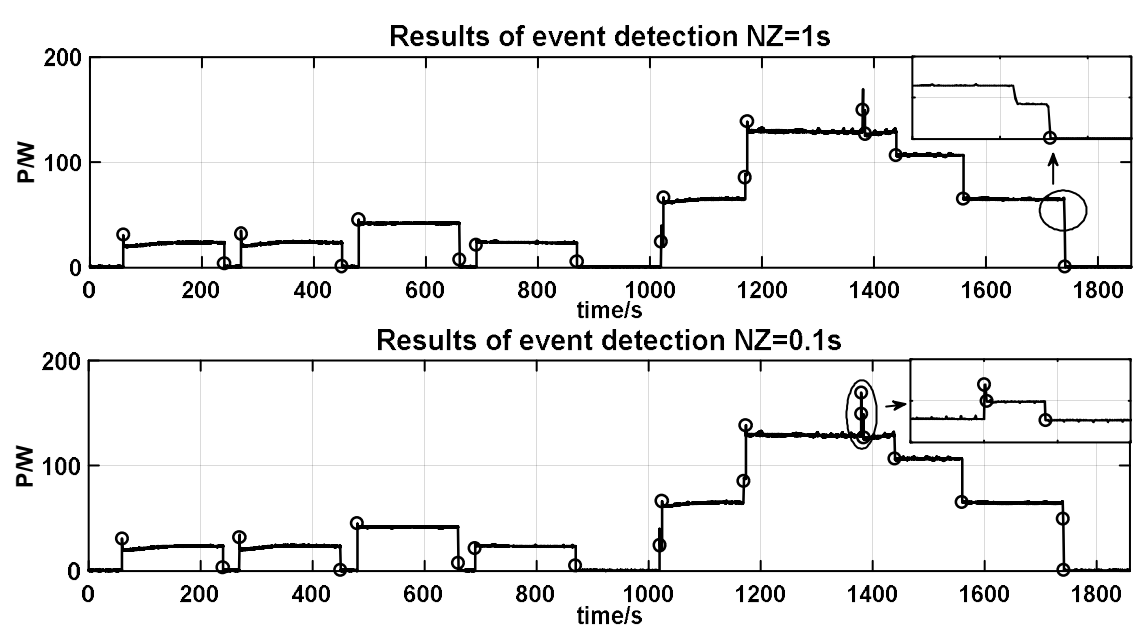

Fig. 24. Event detection results of AWB approach

## C. *Public dataset test in higher frequency*

In this section, we use the BLUED [17] public dataset to test the proposed hybrid approach. The dataset is collected from a real household in Pittsburgh which is designed for event-based NILM approaches. It contains aggregate voltage/current and active power which were sampled at 12 kHz and 60 Hz respectively. The ground-truth events data were collected by using plug-level voltage/current sensors, environmental sensors and sub-circuit sensors.

In this test, we use the one-day data in phase-A as example. In this period, there are 125 recorded events (including 8 simultaneous events on 4 different times) for 12 different appliances (including 4 unknown ones). The algorithm detects the simultaneous events with no time difference as one event. Thus, there are 121 ground-truth events at different times since we consider the simultaneous events as one event. Fig. 25 shows the event detection results of this case.

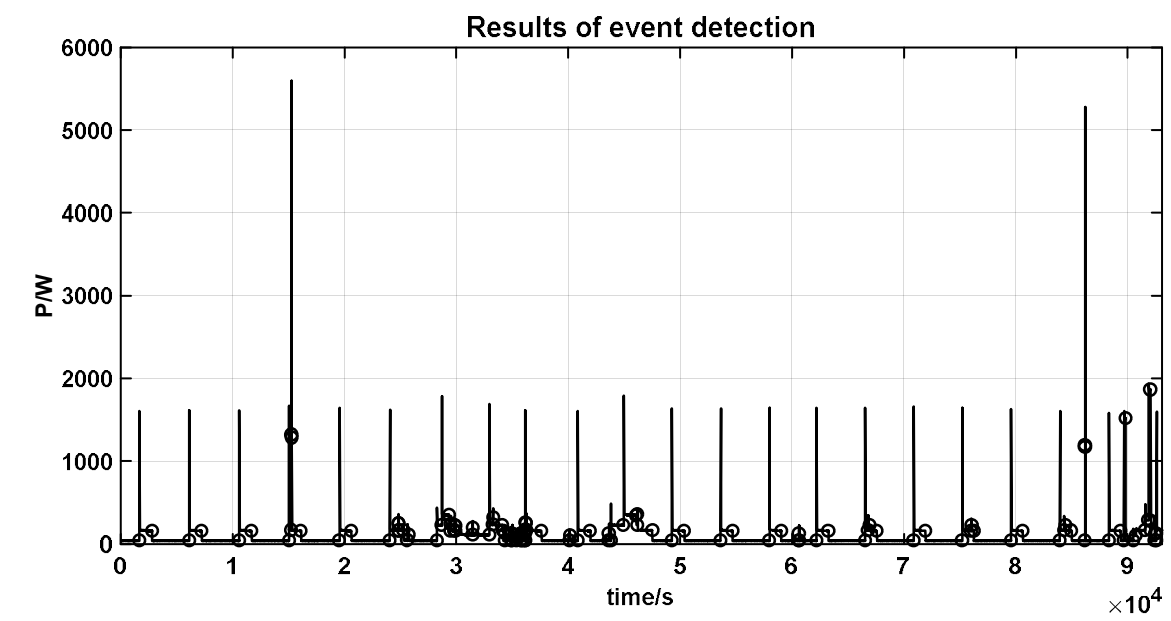

Fig. 25. Event detection results of BLUED

The detection results are evaluated by using true positive rate (TPR=TP/E), false positive rate (FPR=FP/E) and false negative rate (FNR=FN/E), where $E$ is the number of ground-truth events; $TP$, $FP$ and $FN$ are the number of true positive, false positive, and false negative, respectively. Table VI shows the performance of the proposed approach in the above case.

TABLE VI
PERFORMANCE EVALUATION OF EVENT DETECTION ALGORITHM

| TPR | FPR | FNR |
|---|---|---|
| 96.7% | 0.81% | 3.3% |

The number of detected events is 118 which includes 1 false alarm and 4 missing events. These missing events are mainly due to short operation time among multiple appliances. Table VII shows the event detection results after applying each algorithm. The filtering process does not remove false alarms because the involved appliances do not cause much fluctuation.

TABLE VII
EVENTS RESULTS AFTER EACH ALGORITHM

| Algorithm | Events number |
|---|---|
| Base algorithm | 132 |
| Derivative analysis | 118 |
| Filtering | 118 |

Table VIII shows the performance evaluation of the GLR algorithm used in Ref. [9] for the same dataset as comparison. After tuning the parameters to the highest true positive true rate, the algorithm detects 146 events including 29 false alarms and 4 missing events. The false alarms appear mainly because of long-transient period. In this case, the proposed approach achieves the same positive true rate as the algorithm in [9] but with much lower false positive rate which means our approach has the capability of removing false events while maintaining true events.

TABLE VIII
PERFORMANCE EVALUATION OF EVENT DETECTION ALGORITHM IN [9]

| TPR | FPR | FNR |
|---|---|---|
| 96.7% | 24% | 3.3% |

Table IX shows the performance evaluation of the entire one-week data in phase-A of the BLUED [17] public dataset. There are 904 events out of which 30 are simultaneous events. As indicated before, we take simultaneous events as one event which results in 889 true events occurring at different times. In addition, the 94.15% of true positive rate proves the proposed approach can be applied in situations with different sampling rates. The low false positive rate means less false alarms are triggered. Generally, a higher sampling frequency often provides a better resolution and more details, on the other side, it brings more uncertainties to the algorithm. Therefore, the proposed method is robust to sampling frequency in some cases.

TABLE IX
PERFORMANCE EVALUATION OF EVENT DETECTION ALGORITHM

| TPR | FPR | FNR |
|---|---|---|
| 94.15% | 0.79% | 5.85% |

## V. CONCLUSION

In this paper, we propose a hybrid approach for event detection in NILM. It does not require additional information other than the aggregated data collected from a single meter. This approach works without knowing historical data about appliance and houses. In addition, it can improve the detection performance of events which are caused by complex transition period. It also has some abilities to be used in complex situations such as near-simultaneous actions and high fluctuations appliance. Rather than using a single step, the proposed hybrid approach uses one base algorithm and two auxiliary algorithms to detect events and remove false events at the same time. In this process, the base algorithm uses a strict condition to capture as many true events as possible. Each auxiliary method aims to remove certain type of false alarm caused by appliances with long and complex transitions and high fluctuations. Derivative analysis relies on geometrical characters of appliance to find true events among multiple results. It removes the false events on the same appliance transient. The filtering analysis is to remove the wrongly detected events caused by fluctuation of large appliance. In order to prevent true events from being removed by mistake, derivative peaks comparison is introduced to ensure the accuracy of this removal. The auxiliary algorithms remove the detected false events caused by the base algorithm step by step and finally obtain as many true events as possible. The entire approach provides a simple way for event detection without requiring additional information about house or appliance. It also does not need training or an event library. In this paper, we evaluate this algorithm in special cases and public dataset. The results show that it is competitive with other algorithms. Moreover, it does not need to restrict application condition and it has ability to deal with special actions of household appliance such as close-action, complex transition period and small appliance action among large fluctuation working device. We also test the algorithm in real house measurement with different sampling rates to prove its robustness to multiple meters with distinct frequencies.

The proposed approach calculates the mean values of the aggregated data in the base algorithm once the new data are collected from the meter. Then the corresponding derivatives are computed. The filtering analysis recalls the base algorithm when filtering is applied to the aggregated data. Therefore, the computational burden increases along with the increase in sampling rates of the meter. However, this approach does not require advanced calculation or training. Thus, this approach can be applied as an on-line method. The detection lag is decided by the largest window used in the approach which is the time threshold of multiple events (2s in this paper).

Our future work is to apply the proposed hybrid approach to extra electrical parameters such as current, reactive power and power factor to seek for a better performance. Based on the approach, we will employ it in more measurements from different conditions to find a practical way for real house implementation.